\documentstyle [11pt]{article}

\begin{document}

\begin{center}
{\Large \bf The Irreducible Tensor Bases of Exceptional Lie Algebras \ref{1}. 
$G_2$, $F_4$ and $E_6$ }
\end{center}

\vspace{.3cm}

\begin{center}
Dong Ruan${}^{1,2,3}$, Hong-Zhou Sun${}^{1,3}$ and Qi-Zhi Han${}^{4}$
\end{center}

${}^{1}$ Department of Physics, Tsinghua University, Beijing 100084, P.R. China

${}^{2}$ Key Laboratory for Quantum Information and Measurements of MOE, 
Tsinghua University, Beijing 100084, P.R. China

${}^{3}$ Center of Theoretical Nuclear Physics, 
                National Laboratory of Heavy Ion Accelerator, 
                Lanzhou, 730000, P.R. China

${}^{4}$ Department of Physics, Peking University, Beijing 100871, 
P.R. China

\vspace{1cm}

\begin{center}
\begin{minipage}{14cm}
{{\bf Abstract}. 
The irreducible tensor bases of exceptional Lie algebras $G_2$, $F_4$
and $E_6$ are built by grouping their Cartan-Weyl bases according to 
the respective chains $G_2$ $\supset$ SO(3) $\otimes$ SO(3), 
$F_4$ $\supset$ SO(3) $\otimes$ SO(3) $\otimes$ SO(3) $\otimes$ SO(3) 
and 
$E_6$ $\supset$ SO(3) $\otimes$ SO(3) $\otimes$ SO(3) $\otimes$ SO(3).
The explicit commutation relations of the irreducible tensor bases of 
these algebras are given also respectively.
}
\end{minipage}
\end{center}

\vspace{1cm}

{\bf PACS:} \hspace{2mm} 03.65.Fd \hspace{2mm} 03.65.-w  \hspace{2mm} 02.20.Sv

\vfill

\hfill     {Typeset by LaTeX}

\newpage

%%%%%%%%%%%%%%%%%%    Text   %%%%%%%%%%%%%%%%%%%%%%%%%%%

\section{Introduction}
\label{1}
It is well known that the representation theory of Lie groups
has now been established as an invaluable tool in modern physics, especially
in those fields where the symmetry plays an important role, such as atomic,
molecular, nuclear, particle, solid physics and so on.

\par
The complex semisimple Lie algebras were classified completely by 
Cartan \cite{cartan} in his thesis of 1894. Cartan identified four great
classes of Lie algebras, often referred to as the classical Lie algebras, 
$A_n$, $B_n$, $C_n$ and $D_n$, and five exceptional Lie algebras $G_2$, $F_4$,
$E_6$, $E_7$ and $E_8$, where the subscripted integers are the ranks of
the respective algebras. Now the traditional representation theory of Lie
algebra, developed by Cartan, Weyl \cite{weyl}, Chevalley \cite{che} and 
many others, may be found in the numerous standard mathematical textbooks 
(for example, Refs. \cite{hum,var}), in which for Lie algebras there are 
two kinds of bases of very usefulness: the Cartan-Weyl basis and 
the Chevalley basis. 

\par
From the point of view of practical applications to physics, it has been 
proved convenient to have explicit bases for Lie algebras in terms of 
some physical group chains. 
For example, in order to classify the states of electrons in the atomic 
$f$-shell according to the chain SO(7) $\supset$ $G_2$ $\supset$ SO(3), 
Racah \cite{racah} found first that it is possible to build tensor bases of 
SO(7) and $G_2$ by use of the irreducible tensor operators $v^k_q$ of rank $k$
of the three dimensional rotation group SO(3). We now call the kind of tensor
operator realization of Lie algebra the irreducible tensor basis of Lie 
algebra. Later, this idea of Racah was extended to classical Lie algebras 
\cite{judd,jeugt94}, exceptional Lie algebras $F_4$ \cite{jbm,jeugt92,berghe}, $E_6$ 
\cite{berghe,bmj} and $E_7$ \cite{berghe} by considering the chains of these
algebras ending at an SO(3) algebra, and to $F_4$ \cite{mbj} and $E_6$ \cite{bmj}
by the chains of these algebras ending at the direct product of two SO(3) algebras
so that the irreducible tensor bases of $F_4$ and $E_6$ are made up of 
SO(3) $\otimes$ SO(3) irreducible tensor operators 
$v^{k_1\,k_2}_{\:q_1\,q_2}$ of rank $(k_1 k_2)$. 
Furthermore, based upon the irreducible tensor bases, the structural zero of certain
$6j$-coefficients of exceptional Lie algebras has been explained in Racah's spirit. 
However the relationship between the irreducible tensor basis and the standard 
Cartan-Weyl basis is not revealed, and the explicit commutation relations satisfied by
the irreducible tensor bases of these algebras are not given, which are very important
for the problems of irreducible representations. 

\par
In practice the irreducible tensor bases of Lie algebras may be
realized by the other approaches. The alternative useful realization is
based upon a chain of the algebra $G$ under consideration ending at a 
direct product of several SO(3) algebras, i.e., $G$ $\supset$ SO(3) 
$\otimes$ SO(3) $\otimes$ ... $\otimes$ SO(3). Initially, for the classical 
Lie algebras of rank 2 such as $A_2$, $B_2$, $C_2$ and $D_2$, their 
irreducible tensor bases have been build by many authors according to 
the chains $A_2$ $\supset$ SO(3) \cite{s1}, $B_2$ $\supset$ SO(3) 
$\otimes$ SO(3) \cite{he,kpw,s2}, $C_2$ $\supset$ SO(3) $\otimes$ SO(3) 
\cite{ph,bernards} and $D_2$ $\supset$ SO(3) $\otimes$ SO(3) \cite{s3}
respectively. Later, the irreducible tensor bases of the classical 
Lie algebras $A_n$, $B_n$, $C_n$ and $D_n$ of arbitrary rank $n$ 
have been obtained systematically from their respective Cartan-Weyl bases
by Sun and Han \cite{sh1}, and the explicit commutation relations of 
the irreducible tensor bases have been given as well. 
Clearly, the type of irreducible tensor basis of $G$, different from 
Racah's type, are made up of mutually commuting scalar operators, 
mutually commuting angular momentum operators and multi-fold irreducible 
tensor operators of half-odd integral ranks.
The purpose of the present paper is to generalize the method applied in
\cite{sh1} to construct the irreducible tensor bases of five exceptional 
Lie algebras. Owing to the fact that the exceptional Lie algebras 
$F_4$, $E_6$, $E_7$ and $E_8$ contain the classical Lie algebras 
$B_4$, $A_5$, $A_7$ and $D_8$ 
% (Their irreducible tensor basis may be %found in Ref. \cite{sh1}) 
as subalgebras respectively, therefore, 
this work constructing the irreducible tensor bases of 
$F_4$, $E_6$, $E_7$ and $E_8$ is to group their remaining generators 
corresponding to the extra roots by the similar approach used in \cite{sh1}
to yield the extra scalar operators, angular momentum operators and multi-fold
irreducible tensor operators.
In this paper, only the irreducible tensor bases of $G_2$, $F_4$ and $E_6$ 
are constructed, whereas those of $E_7$ and $E_8$ will be discussed in a
subsequent paper. 

\par
This paper is arranged as follows. In Section \ref{2}, the basic
definitions to be employed such as angular momentum operator, scalar 
operator and (multi-fold) irreducible tensor operator and 
the notation of Cartan-Weyl basis of Lie algebra are reviewed briefly.
In Sections \ref{3}-\ref{5}, the irreducible tensor bases of $G_2$, 
$F_4$ and $E_6$ are constructed respectively, and the explicit 
commutation relations satisfied by the irreducible tensor 
bases are calculated also. 
A simple conclusion is given in the final section. 

\section{ Definitions and notation }
\label{2}
1) {\bf Angular momentum operator} 
\par
{\bf J}(1), {\bf J}(2), ..., {\bf J}$(n)$ are $n$ mutually
commuting angular momentum operators, if they satisfy 
\begin{eqnarray}
\begin{array}{l}
\left[ J_0(i)\, , \hspace{2mm}J_{\pm 1}(i) \right] = \pm J_{\pm 1}(i), \\ 
\left[ J_{+1}(i)\, , \hspace{2mm}J_{-1}(i) \right] = - J_0(i); \\ 
\left[{\bf J}(i)\, , \hspace{2mm} {\bf J}(j) \right] = 0, 
\hspace{3mm} i \not= j ,
\end{array}
\label{angular-d}
\end{eqnarray}
where $J_{+1}(i)$, $J_0(i)$ and $J_{-1}(i)$ are the spherical components 
of the $i$th angular momentum operator {\bf J}$(i)$, or, in mathematical
language, are the infinitesimal generators of a SO(3) group. 

{\noindent
2) {\bf Scalar operator }  
}
\par
$A(1)$, $A(2)$, ..., $A(n)$ are $n$ mutually commuting scalar operators, 
if they commute amongst themselves and with all angular momentum operators,
i.e., 
\begin{eqnarray}
\begin{array}{l}
\left[ {\bf J}(i)\, , \hspace{2mm} A(j) \right] = 0 , \\ 
\left[ A(i)\, , \hspace{2mm} A(j) \right] = 0 .
\end{array}
\label{scalar-d}
\end{eqnarray}

{\noindent
3) {\bf Irreducible tensor operator} \cite{ra,edmonds,wy,bl}  
}
\par
${\bf U}^{r}(i)$ is a irreducible tensor operator of rank $r$ with respect to
the $i$th angular momentum operator (or SO(3)), if its $2r+1$ components 
$U^{r}_{p}(i)$, together with the angular momentum operators, satisfy 
\begin{eqnarray}
\begin{array}{l}
\left[ J_0(i)\, , \hspace{2mm} U^{r}_{p}(i) \right] = p\,U^{r}_{p}(i), \\ 
\left[ J_{\pm 1}(i)\, , \hspace{2mm} U^{r}_{p}(i) \right] 
        = C_{\pm}(r p)\,U^{\;r}_{p\pm 1}(i);                        \\ 
\left[ {\bf J}(j)\, , \hspace{2mm} U^{r}_{p}(i) \right] = 0 , 
\hspace{3mm} 
j\not= i,
\end{array}
\label{tensr-d}
\end{eqnarray}
where $r$ may take nonnegative integers or half-odd integers and for a fixed 
$r$, the component label $p$ may take $-r$, $-r+1$, ..., $r$, and 
\[
C_{\pm}(r p) = \mp \sqrt{{\frac{1}{2}} (r\mp p)(r\pm p + 1)} . 
\]

{\noindent
4) {\bf Multifold irreducible tensor operator}  
}
\par
${\bf U}^{r_1, ..., r_m }(i_1, ..., i_m)$ is a $m$-fold
irreducible tensor operator of rank $(r_1$, ..., $r_m)$ with respect to
SO(3) $\otimes$ SO(3) $\otimes$ ... $\otimes$ SO(3) (i.e., direct
product of $m$ SO(3) groups), if its 
$(2 r_1 +1)$ $\times$ $(2 r_2 +1)$ $\times$...$\times$ $(2r_m + 1)$ components 
$U^{r_1, ..., r_m}_{\,p_1, ..., p_m}(i_1, ..., i_m)$, together with the angular
momentum operators, satisfy the following relations 
\begin{eqnarray}
\begin{array}{l}
\left[ J_0(i_{\alpha})\, , \hspace{2mm} 
       U^{r_1 ... r_{\alpha}... r_m }
        _{\,p_1 ... p_{\alpha}... p_m}(i_1 ...i_{\alpha}... i_m)
\right]
 = p_{\alpha} \, 
       U^{r_1 ... r_{\alpha}... r_m }
        _{\,p_1 ... p_{\alpha}... p_m}(i_1 ...i_{\alpha}... i_m),          \\ 
\left[ J_{\pm 1}(i_{\alpha})\, , \hspace{2mm} 
       U^{r_1 ... r_{\alpha}... r_m }
        _{\,p_1 ... p_{\alpha}... p_m}(i_1 ...i_{\alpha}... i_m)
\right] 
  = C_{\pm}(r_{\alpha}\,p_{\alpha}) \,
       U^{r_1 ... \;\; r_{\alpha}\;\;...\, r_m }
        _{\,p_1 ... p_{\alpha}\pm1... p_m}(i_1 ...i_{\alpha}... i_m);      \\ 
\left[ {\bf J}(i_{\beta})\, , \hspace{2mm} 
       U^{r_1 ... r_m }_{\,p_1 ...\,p_m} (i_1, ..., i_m) 
\right] = 0, 
\hspace{5mm} 
i_{\beta} \not= i_1,\;i_{\beta} \not= i_2,\, ..., \, i_{\beta} \not= i_m,
\end{array}
\label{mtensr-d}
\end{eqnarray}
where 
\[
C_{\pm}(r_{\alpha}\, p_{\alpha})  = \mp \sqrt{{\frac{1}{2}} (r_{\alpha} \mp
p_{\alpha}) (r_{\alpha} \pm p_{\alpha} + 1)} . 
\]
Since a $m$-fold irreducible tensor operator is in fact a direct product 
of $m$ mutually independent irreducible tensor operators, so the restriction 
among $r_i$'s in rank $(r_1$, ..., $r_m)$ does not exit so that, 
similar to the definition (3), any $r_i$ ($i=1$, $2$, ..., $m$) may 
take nonnegative integers or half-odd integers and for a fixed $r_i$, 
the corresponding component label $p_i$ may take $-r_i$, $-r_i+1$, ..., 
$r_i$.

\par
We see from the definitions (2)-(4) that only the concept of 
irreducible tensor operator is basic, whereas a scalar operator is 
a special irreducible tensor operator of rank 0 and the concept of 
multi-fold irreducible tensor operator is a natural extension of 
the definition of irreducible tensor operator.

\par
{\it Notation}: Let $\{ H_1$, $H_2$, ..., $H_n$; $E_{\pm \alpha}
$, $\alpha \in \sum^+ \}$ be the Cartan-Weyl basis of some exceptional Lie
algebra of rank $n$ \cite{hum,ra,wy}, where $\sum^+$ is its positive 
root system.  
In this paper, we will use the simple notation, for example, 
when $\alpha$ $=$ $e_i - e_j$, the corresponding generators 
$E_{\pm (e_i - e_j)}$ are replaced by $E_{\pm (i-j)}$.

\section{The irreducible tensor basis of $G_2$}
\label{3}
In order to see clearly how to construct the irreducible tensor basis 
from the corresponding Cartan-Weyl basis, let us begin with 
the simplest exceptional Lie algebra $G_2$.

\par
As is known, $G_2$ has twelve nonnull roots \cite{ra,wy}
$$
  e_i - e_j, \hspace{6mm} \pm (e_i + e_j)\mp 2e_k, \hspace{6mm} 
  1\le i<j <k \le 3,  
$$
with the normalization constant $ K=\sqrt{24} $.

\par
In terms of the symmetries and identities satisfied by
the structure constants \cite{ra,wy}, we take
the structure constants of the Cartan-Weyl basis of $G_2$ as
$$
  N_{61}= N_{64}= N_{42}= N_{15}
        = {1\over 2}\,\sqrt{1 \over 2}, \hspace{5mm}  
  N_{61}= \sqrt{1 \over 6}. 
$$
Then we may let
\begin{eqnarray*}
\begin{array}{l}   
  J_0(1) = 2\,\sqrt3 \, H_1,  \\
  J_{\pm 1}(1) = \mp \, 2\,\sqrt3 \, E_{\pm 3} ;          \\  
  J_0(2) = 2\, H_2,           \\
  J_{\pm 1}(2) = \mp \, 2\, E_{\pm 6},
\end{array}
\label{g2-angular}
\end{eqnarray*}
and put $U^{{1\over 2}{3\over 2}}_{\,p\,q}(12)\equiv
U_{\,p\,q}(12)$ as
\begin{center}
{\tabcolsep 12 pt
\doublerulesep 0pt
\begin{tabular}{rllll} \hline
   $p\backslash q$ 
       &  ${3\over 2}$   &  ${1\over 2}$  
       &  $-{1\over 2}$  & $-{3\over 2}$ \\  \hline
 $ \hspace{2.2mm}{1\over 2}$ &  $ 2\,\sqrt3 \, E_{5}$ 
       &  $ 2\,\sqrt3 \, E_{4} $  
       &  $ 2\,\sqrt3 \, E_{2} $  &  $ 2\,\sqrt3 \, E_{1} $     \\ 
 $ -{1\over 2}$
       &  $ -2\,\sqrt3 \, E_{-1}$ &  $ 2\,\sqrt3 \, E_{-2} $
       &  $ -2\,\sqrt3 \, E_{-4} $  &  $ 2\,\sqrt3 \, E_{-5} $  \\
\hline         
\end{tabular} }
\end{center}
The number of the above operators is $3+3+2\times 4 =14$, it is 
equal to the order of $G_2$. Hence, these operators form
the irreducible tensor basis of $G_2$. The root diagram 
corresponding to the irreducible tensor basis of $G_2$ is given
in Figure 1.
                
\par
It is easy to find that the irreducible tensor operator 
${\bf U}^{{1\over 2}{3\over 2}}(12)$ and its Hermitian conjugate 
$({\bf U}^{{1\over 2}{3\over 2}}(12))^{\dagger}$ 
satisfy the following relation
\begin{eqnarray*}
U_{-p\,-q}(12) = (-)^{p+q} \, U_{p\,q}^{\dagger}(12).
\end{eqnarray*}

\par
By direct calculations, we can obtain the commutation relations
satisfied by the irreducible tensor basis of $G_2$: 
\par
1)  {\bf J}(1) and {\bf J}(2) 
   are two mutually commuting angular momentum operators, 
   and satisfy the commutation relations (\ref{angular-d}).

\par
2) ${\bf U}^{{1\over 2}{3\over 2}}(12)$ is a 2-fold irreducible tensor
operator, hence it, together with {\bf J}(1) and {\bf J}(2), satisfies 
the commutation relations (\ref{mtensr-d}).

3) The nonzero commutation relations between eight components 
$U_{\,p\,q}(12)$ 
($p= -{1\over 2}$, ${1\over 2}$, and $q= -{3\over 2}$, $- {1\over 2}$, ${1\over 2}$, 
${3\over 2}$)
of ${\bf U}^{{1\over 2}{3\over 2}}(12)$ are given in coupled irreducible tensor 
forms, which are more compact and symmetric than usual Lie bracket forms, as follows: 
\begin{eqnarray}
\begin{array}{l}
    \left( {\bf U}^{{1\over 2}{3\over 2}}(12) \, {\bf U}^{{1\over 2}{3\over 2}}(12)
    \right)^{1\,0}_{\mu\,0} 
      =  \sqrt{9\over 2}\, J_{\mu}(1) ,  \cr
    \left( {\bf U}^{{1\over 2}{3\over 2}}(12) \, {\bf U}^{{1\over 2}{3\over 2}}(12) 
    \right)^{0\,1}_{0\,\mu} 
      =  \sqrt{5\over 2}\, J_{\mu}(2) ,  \cr 
         \mu = -1,\,0,\,1.            
\end{array}
\end{eqnarray}
Here (and afterwards) we have used the definition of coupled irreducible tensor operator, 
\cite{fano} for example, for two 2-fold irreducible tensor operators ${\bf U}^{k_1 l_1}(ij)$ 
and ${\bf U}^{k_2 l_2}(ij)$,
which correspond to the common angular momenta ${\bf J}(i)$ and ${\bf J}(j)$ 
(i.e., labels $k_1$ and $k_2$ correspond to ${\bf J}(i)$, and labels $l_1$ and $l_2$ 
to ${\bf J}(j)$), thus we may couple ${\bf U}^{k_1 l_1}(ij)$ and ${\bf U}^{k_2 l_2}(ij)$
to produce a new 2-fold irreducible tensor operator indicated as
$( {\bf U}^{k_1 l_1}(ij) \, {\bf U}^{k_2 l_2}(ij) )^{k\,l}$, 
whose components are 
\begin{eqnarray}
\begin{array}{rl}
    \left( {\bf U}^{k_1 l_1}(ij) \, {\bf U}^{k_2 l_2}(ij) \right)^{k\,l}_{pq} =
      & \sum_{p_1\,p_2\,q_1\,q_2}\: \langle k_1 \, p_1 \, k_2 \, p_2\,
        \left| \right. k \, p \rangle \,
        \langle l_1 \, q_1 \, l_2 \, q_2\,
        \left| \right. l \, q \rangle                       \cr
   {} &  \times U^{k_1 l_1}_{\,p_1 \,q_1}(ij) \, U^{k_2 l_2}_{\,p_2 \,q_2}(ij) ,
\end{array}
\label{define}
\end{eqnarray}
where symbol
$
 \langle k_1 \, p_1 \, k_2 \,p_2\, \left| \right. k \, p \rangle
$
is the usual Clebsch-Gordon coefficient of SO(3), \cite{edmonds,bl} for the given
$k_1$ and $k_2$, $k$ may take $|k_1- k_2|$, $|k_1- k_2|+1$, ..., $k_1 + k_2$. 
Especially, if $k_1 = k_2$, and when $k$ takes 0, then 
$( {\bf U}^{k_1 l_1}(ij) \, {\bf U}^{k_2 l_2}(ij) )^{0\,l}$ is just a 
(1-fold) irreducible tensor operator of rank $l$. If $k_1 = k_2$ and $l_1 = l_2$, 
and when $k=l=0$, then  
$( {\bf U}^{k_1 l_1}(ij) \, {\bf U}^{k_2 l_2}(ij) )^{00}$ is just a 
scalar operator. 

\par
It is very obvious from the above commutation relations that 
the Cartan generators of $G_2$ in the scheme of irreducible tensor basis are 
$\{ J_0(1)$, $J_0(2) \}$.

\section{The irreducible tensor basis of $F_4$}
\par
It is known \cite{ra,wy} that $F_4$ contains $B_4$ as a subalgebra, 
hence, all nonnull roots of $F_4$  include those of $B_4$
\[  \pm e_i, \hspace{6mm} \pm e_i \pm e_j, \hspace{6mm} i<j,  
    \hspace{6mm}  i,\,j=1,\,2,\,3,\,4
\]
and the extra roots
\[ {1 \over 2} (\pm e_1 \pm e_2 \pm e_3 \pm e_4).     \]

\par
For convenience, let
\begin{eqnarray*}
    \begin{array}{l}  
     \alpha   = {1\over 2}\,\{  (e_1+ e_2) \pm (e_3+ e_4)\},   \\
     \beta    = {1\over 2}\,\{  (e_1+ e_2) \pm (e_4- e_3)\},   \\
     \gamma   = {1\over 2}\,\{  (e_2- e_1) \pm (e_3+ e_4)\},   \\
     \epsilon = {1\over 2}\,\{  (e_2- e_1) \pm (e_4- e_3)\},
  \end{array}
\end{eqnarray*}
with
\begin{eqnarray*} 
    y_1 = \{ (\cdots) + (\cdots)\}  , \hspace{5mm} 
    y_2 = \{ (\cdots) - (\cdots)\} ,  
\end{eqnarray*}
where $y$ may take $\alpha$, $\beta$, $\gamma$, $\epsilon$.

\par
The irreducible tensor basis of $B_4$ has been given in Ref. \cite{sh1}.
Thus in terms of the symmetries and identities satisfied by the structure
constants \cite{ra,wy}, we take the structure constants of 
the Cartan-Weyl basis of $F_4$ as
\[ N_{ij}= N_{j,\,i-j}= N_{i-k,\,j+k}= N_{i+k,\,j-k}
         = N_{j+k,\,i-j}= N_{j-k,\,i-j}= -{1\over K} ,
\]
\[  i<j <k \leq 4; \hspace{10mm} N_{xy} = {G_{xy}\over K},
\]
where $N_{ij}$, $N_{j,\,i-j}$, ..., $N_{j-k,\,i-j}$ are the structure 
constants of $B_4$ and $G_{xy}$ is given in Table 1. 

\par
Now we may let
\begin{eqnarray*}
  \begin{array}{l}
    J_0(i^{\prime}) = {K\over 2}(H_{i^{\prime}}
                                      + H_{i_1^{\prime}}) ,         \\
    J_{\pm 1}(i^{\prime}) = \pm {K\over \sqrt{2}} 
                                 E_{\pm(i^{\prime} +i_1^{\prime})}; \\
    J_0(i_1^{\prime}) = {K\over 2} (-H_{i^{\prime}}
                                         +H_{i_1^{\prime}}) ,        \\
    J_{\pm 1}(i_1^{\prime}) = \pm {K\over \sqrt{2}} 
                                 E_{\pm(-i^{\prime} +i_1^{\prime})} , \\
    i^{\prime}=1, 3,  \hspace{3mm} 
    i_1^{\prime}=i^{\prime} + 1,
\end{array}
\end{eqnarray*}          
and put $U_{\;p\:q}^{{1\over 2}{1 \over 2}}(ij) \equiv U_{p\: q}(ij)$
as
\begin{center}
{\renewcommand\arraystretch{1.1}
\tabcolsep 10mm
\doublerulesep 0pt
\begin{tabular}{ccc} \hline
   $ p \backslash q $ & $ {1\over 2} $   & $-{1\over 2} $    \\ \hline
  $ {1\over 2} $ & $-\frac{K}{\sqrt{2}}E_{i_1^{\prime}} $
                 & $\frac{K}{\sqrt{2}}E_{i^{\prime}}    $   \\
  $ -{1\over 2}$ & $\frac{K}{\sqrt{2}}E_{-i^{\prime}}   $
                 & $\frac{K}{\sqrt{2}}E_{-i_1^{\prime}} $   \\ \hline
\end{tabular} }
\end{center}
for $i\,j = i^{\prime}\, i_1^{\prime}$, or
\par 
\begin{center}
{\renewcommand\arraystretch{1.1}
\tabcolsep 10mm
\doublerulesep 0pt
\begin{tabular}{ccc} \hline
   $ p \backslash q $ & $ {1\over 2} $   & $-{1\over 2} $    \\ \hline
  $ {1\over 2}$  & $-\frac{K}{\sqrt{2}}E_{y_1} $
                 & $\frac{K}{\sqrt{2}}E_{y_2}  $       \\
  $ -{1\over 2}$ & $\frac{K}{\sqrt{2}}E_{-y_2} $
                 & $\frac{K}{\sqrt{2}}E_{-y_1} $  \\ \hline
\end{tabular}  }
\end{center}
for $i\,j \not= i^{\prime}\, i_1^{\prime}$, where 
\begin{eqnarray*}
     i\, j = \left\{
  	   \begin{array}{lll}
                1\,3 ,  &  \mbox{when} \hspace{2mm} y= \alpha; \cr 
                1\,4 ,  &  \mbox{when} \hspace{2mm} y= \beta; \cr
                2\,3 ,  &  \mbox{when} \hspace{2mm} y= \gamma; \cr 
                2\,4 ,  &  \mbox{when} \hspace{2mm} y= \epsilon,
           \end{array}   
             \right.
\end{eqnarray*}
and  put
$U_{\;p\:q\:p'\:q'}^{{1\over 2}{1 \over 2}\,{1 \over 2}\:{1 \over 2}}(1234)
\equiv U_{p\: q\: p'\: q'}(1234)$ as
\begin{center}
{\renewcommand\arraystretch{1.1}
\tabcolsep 5mm
\doublerulesep 0pt
\begin{tabular}{rllll} \hline
        $p\,q \backslash p' \, q' $
     &  ${1\over 2}\; {1\over 2}  $ 
     &  ${1\over 2}\; -{1\over 2} $
     &  $-{1\over 2}\; {1\over 2} $ 
     & $-{1\over 2}\; -{1\over 2}$    \\ \hline
${1\over 2}\hspace{5mm}{1\over 2} \hspace{5mm} $   
                             & $-\frac{K}{\sqrt{2}}E_{2+4} $
                             & $ \frac{K}{\sqrt{2}}E_{2+3} $   
                             & $-\frac{K}{\sqrt{2}}E_{2-3} $
                             & $-\frac{K}{\sqrt{2}}E_{2-4} $ \\
${1\over 2}\; -\!{1\over 2} \hspace{5mm} $ 
                             & $ \frac{K}{\sqrt{2}}E_{1+4} $
                             & $-\frac{K}{\sqrt{2}}E_{1+3} $   
                             & $ \frac{K}{\sqrt{2}}E_{1-3} $
                             & $ \frac{K}{\sqrt{2}}E_{1-4} $ \\
$-{1\over 2}\hspace{5mm}{1\over 2} \hspace{5mm} $ 
                             & $ \frac{K}{\sqrt{2}}E_{-1+4} $
                             & $-\frac{K}{\sqrt{2}}E_{-1+3} $   
                             & $ \frac{K}{\sqrt{2}}E_{-1-3} $
                             & $ \frac{K}{\sqrt{2}}E_{-1-4} $ \\
$-{1\over 2}\;-\!{1\over 2} \hspace{5mm} $
                             & $ \frac{K}{\sqrt{2}}E_{-2+4} $
                             & $-\frac{K}{\sqrt{2}}E_{-2+3} $   
                             & $ \frac{K}{\sqrt{2}}E_{-2-3} $
                             & $ \frac{K}{\sqrt{2}}E_{-2-4} $ 
\\ \hline  
\end{tabular} }
\end{center}
It is easy to get
\begin{eqnarray*} 
U_{-p\,-q}(ij)  = (-)^{p+q}\, U_{p\,q}^{\dagger}(ij),
\end{eqnarray*}
\begin{eqnarray*} 
U_{-p\,-q\,-p'\,-q'}(1234)  = 
(-)^{1+p+q+p'+q'}\, U_{p\,q\,p'\,q'}^{\dagger}(1234).
\end{eqnarray*}

\par
The number of the above operators is
$ 3\times 4 + 4\times 2 + 4\times 4 + 16 = 52$, it is 
equal to the order of $F_4$. Hence, these operators form
the irreducible tensor basis of $F_4$. 
                   
\par
By direct calculations, we can obtain the commutation relations satisfied by
the irreducible tensor basis of $F_4$: 
\par
1) {\bf J}(1), {\bf J}(2), {\bf J}(3) and {\bf J}(4)
   are the mutually commuting angular momentum operators, 
   and satisfy the commutation relations (\ref{angular-d}).

\par
2) ${\bf U}^{{1\over 2}{1\over 2}}(ij)$ ($i=1$, 2; $j=3$, 4) are 
the 2-fold irreducible tensor operators, hence they, together with 
${\bf J}(i)$ and  ${\bf J}(j)$, satisfy the commutation relations 
(\ref{mtensr-d}).

\par
3) ${\bf U}^{{1\over 2}{1\over 2}{1\over 2}{1\over 2}}(1234)$ is
a 4-fold irreducible tensor operator, hence it, together with 
{\bf J}(1), {\bf J}(2), {\bf J}(3) and {\bf J}(4), satisfies 
the commutation relations (\ref{mtensr-d}).

\par
4) The nonzero commutation relations satisfied by the components of 
${\bf U}^{{1\over 2}{1\over 2}}(ij)$ and
${\bf U}^{{1\over 2}{1\over 2}{1\over 2}{1\over 2}}(ijkl)$ read:
\begin{eqnarray}
 \begin{array}{l}
  [\, U_{p\,q}(12),\; U_{p'\,q'}(34)\, ] =  \sqrt{1 \over 2}\,  
	U_{p\,q\,p'\,q'}(1234), \cr
  [\, U_{p\,q}(13),\; U_{p'\,q'}(24)\, ] =  - \sqrt{1 \over 2}\,  
	U_{p\,p'\,q\,q'}(1234), \cr
  [\, U_{p\,q}(14),\; U_{p'\,q'}(23)\,] =  - \sqrt{1 \over 2}\,  
	U_{p\,p'\,q'\,q}(1234); \cr
  \left( {\bf U}^{{1\over 2}{1\over 2}}(ij)\, {\bf U}^{{1\over 2}{1\over 2}}(ij)
  \right)^{10}_{\mu 0} = -{1\over 2} J_{\mu}(i),               \\
  \left( {\bf U}^{{1\over 2}{1\over 2}}(ij)\, {\bf U}^{{1\over 2}{1\over 2}}(ij)\
  \right)^{01}_{0\mu } = -{1\over 2} J_{\mu}(j),  \hspace{5mm}                     
   \mu = -1,\,0,\,1,                  \\
  \left\{ {\bf U}^{{1\over 2}{1\over 2}}(ij) \, {\bf U}^{{1\over 2}{1\over 2}}(jk) 
  \right\}^{{1\over 2}0{1\over 2}}_{\,p\:0\:q} 
           = (-)^{x+1} \sqrt{1\over 2}\, 
             U^{{1\over 2}{1\over 2}}_{\,p\,q}(ik),  \\
   x= \mbox{min}(i,j,k);                    \\
  \left[ {\bf U}^{{1\over 2}{1\over 2}}(ij) \, 
         {\bf U}^{{1\over 2}{1\over 2}{1\over 2}{1\over 2}}(ijkl) 
  \right]  ^{00{1\over 2}{1\over 2}}_{00\,p\,q} 
           = -\sqrt{2}\, U^{{1\over 2}{1\over 2}}_{\,p\,q}(kl), \\
  \left[ {\bf U}^{{1\over 2}{1\over 2}}(ik) \,
	    {\bf U}^{{1\over 2}{1\over 2}{1\over 2}{1\over 2}}(ijkl) 
  \right]  ^{0{1\over 2}0{1\over 2}}_{0\,p\,0q} 
           = (-)^{\, i+1} \sqrt{2}\, 
                U^{{1\over 2}{1\over 2}}_{\,p\,q}(jl);    \\
   \left( {\bf U}^{{1\over 2}{1\over 2}{1\over 2}{1\over 2}}(ijkl) \,
          {\bf U}^{{1\over 2}{1\over 2}{1\over 2}{1\over 2}}(ijkl) 
   \right)^{1000}_{\mu 000} = - J_{\mu}(i),               \\
   \left( {\bf U}^{{1\over 2}{1\over 2}{1\over 2}{1\over 2}}(ijkl)\,
          {\bf U}^{{1\over 2}{1\over 2}{1\over 2}{1\over 2}}(ijkl) 
   \right)^{0100}_{0\mu 00} = - J_{\mu}(j),              \\
   \left( {\bf U}^{{1\over 2}{1\over 2}{1\over 2}{1\over 2}}(ijkl)\,
          {\bf U}^{{1\over 2}{1\over 2}{1\over 2}{1\over 2}}(ijkl) 
   \right)^{0010}_{00\mu 0} = - J_{\mu}(k),              \\
   \left( {\bf U}^{{1\over 2}{1\over 2}{1\over 2}{1\over 2}}(ijkl)\,
          {\bf U}^{{1\over 2}{1\over 2}{1\over 2}{1\over 2}}(ijkl) 
    \right)^{0001}_{000\mu} = - J_{\mu}(l).  
\end{array}
\label{f4-cr}
\end{eqnarray}
Here we have used the following two convenient notations
\begin{eqnarray}
\begin{array}{c}
     \left\{ {\bf X}^{\eta_1} \, {\bf Y}^{\eta_2} \right\}^{\eta}_{\zeta}
   \equiv 
     \left( {\bf X}^{\eta_1} {\bf Y}^{\eta_2} \right)^{\eta}_{\zeta}
      + \left( {\bf Y}^{\eta_2} {\bf X}^{\eta_1} \right)^{\eta}_{\zeta},  \cr
     \left[ {\bf X}^{\eta_1} \, {\bf Y}^{\eta_2} \right]^{\eta}_{\zeta}
   \equiv 
     \left( {\bf X}^{\eta_1} {\bf Y}^{\eta_2} \right)^{\eta}_{\zeta} 
      - \left({\bf Y}^{\eta_2} {\bf X}^{\eta_1} \right)^{\eta}_{\zeta} ,  
\end{array}
\end{eqnarray}
where $\left( {\bf X}^{\eta_1} {\bf Y}^{\eta_2} \right)^{\eta}_{\zeta}$ is 
a coupled irreducible tensor operator of rank $\eta$ (see Eq. (\ref{define})) 
build from two irreducible tensor operators ${\bf X}$ of rank $\eta_1$ and 
${\bf Y}$ of rank $\eta_2$. 
We note that the former three commutation relations in Eq. (\ref{f4-cr}) are 
written in the usual Lie bracket forms since the two irreducible tensor 
operators in Lie brackets do not correspond to the common angular momentum.

\par
We can find easily from the above commutation relations
that the Cartan generators of $F_4$ in the irreducible tensor basis are
$\{ J_0(1)$, $J_0(2)$, $J_0(3)$, $J_0(4) \}$.

\section{The irreducible tensor basis of $E_6$}
\label{5}
\par
It is known \cite{ra,wy} that $E_6$ contains $A_5$ as a subalgebra, 
hence, all nonnull roots of $E_6$  include those of $A_5$
\[ e_i - e_j, \hspace{6mm} i\not= j, \hspace{6mm} i,\,j= 1,\,2,...,\, 6,
\]
and the extra roots
\[ \pm \sqrt{2}\, e_7, \hspace{5mm}  
   {1 \over 2} (\pm e_1 \pm e_2 \pm e_3 \pm e_4 \pm e_5 \pm e_6)
   \pm {1\over \sqrt{2}}\, e_7 ,
\]
where three positive sign and three negative sign are taken
in the above parentheses. The normalization constant of root vectors is 
$K=\sqrt{144}$.

\par
For convenience, let
\begin{eqnarray*}
  \begin{array}{l}  
     \alpha   = {1\over 2}\, \{ (e_1+ e_2- e_3- e_4) \pm (e_6- e_5)
                           \pm \sqrt{2}\, e_7 \},   \\
     \beta    = 
     {1\over 2}\,\{ (e_1+ e_2- e_5- e_6) \pm (e_4- e_3)
                           \pm \sqrt{2}\, e_7 \},   \\
     \epsilon = 
     {1\over 2}\,\{ (e_3+ e_4- e_5- e_6) \pm (e_2- e_1)
                           \pm \sqrt{2}\, e_7 \},   \\
     \lambda   = 
     {1\over 2}\,\{ \pm (e_2- e_1)\pm (e_4- e_3) \pm (e_6- e_5)
                           \pm \sqrt{2}\, e_7 \},   \\
  \end{array}
\end{eqnarray*}
with
\begin{eqnarray*}
  \begin{array}{rl}
   y_1 = \{ (\cdots) + (\cdots) + (\cdots)\}  ,         &
   y_2 = \{ (\cdots) + (\cdots) - (\cdots)\}  , \\     
   y_3 = \{ (\cdots) - (\cdots) + (\cdots)\}  ,         &
   y_4 = \{ (\cdots) - (\cdots) - (\cdots)\}  ;  \\
   x_1 = \{ + (\cdots)+ (\cdots)+ (\cdots)+ (\cdots)\}  , &     
   x_2 = \{ + (\cdots)+ (\cdots)+ (\cdots)- (\cdots)\}  , \\    
   x_3 = \{ + (\cdots)+ (\cdots)- (\cdots)+ (\cdots)\}  , &     
   x_4 = \{ + (\cdots)+ (\cdots)- (\cdots)- (\cdots)\}  , \\     
   x_5 = \{ + (\cdots)- (\cdots)+ (\cdots)+ (\cdots)\}  , &     
   x_6 = \{ + (\cdots)- (\cdots)+ (\cdots)- (\cdots)\}  , \\     
   x_7 = \{ + (\cdots)- (\cdots)- (\cdots)+ (\cdots)\}  , &     
   x_8 = \{ + (\cdots)- (\cdots)- (\cdots)- (\cdots)\}  ,      
  \end{array}
\end{eqnarray*}
where
$  y = \alpha, \, \beta, \,  \epsilon $ and 
$  x= \lambda $.

\par
The irreducible tensor basis of $A_5$ has been given in Ref. \cite{sh1}.
Thus in terms of the symmetries and identities satisfied by the structure
constants \cite{ra,wy}, we take the structure constants of 
the Cartan-Weyl basis of $E_6$ as
\[ N_{i-j,\,j-k}= {1\over K} , \hspace{6mm} 
   i<j <k \leq 6 ; \hspace{6mm} N_{xy} = {S_{xy}\over K},
\]
where $N_{i-j,\,j-k}$ is the structure constant of $A_5$ and 
$S_{xy}$ is given in Table 2.

\par
Now we may let 
\begin{eqnarray*}
\begin{array}{l}
    J_0(i_1) = {K\over 2}\,(-H_i + H_{i_1}) ,         \\
    J_{\pm 1}(i_1) = \pm {K\over \sqrt{2}} \,
                                 E_{\pm(-i + i_1)},            \\
    i_1 =i + 1, \hspace{6mm} i =1,\, 3,\,5;                    \\
    A(i)= K(H_i+ H_{i_1}) ,                                    \\   
    \sum\limits_{i}A(i) =0;                                     \\
    J_0(8) = {K\over \sqrt{2}}\, H_7 ,                 \\         
    J_{\pm 1}(8) = \pm {K\over \sqrt{2}} \,E_{\pm 7} , 
\end{array}
\end{eqnarray*}    
where $E_{\pm 7}$ are the generators corresponding to
the roots $\pm \sqrt{2}\,e_{{}_{7}}$,
and put
$V^{\hspace{1mm} {1\over 2} \hspace{3.5mm} {1\over 2}}_{1\,p\,-1\,q}
$
$(i\,i_1\,j\,j_1)$ $\equiv$ $V_{\,p\,q}(i_1\,j_1)$
and $W^{\hspace{3mm} {1\over 2} \hspace{2.25mm} {1\over 2}}_{-1\,p\,1\,q}
(i\,i_1\,j\,j_1)$ $\equiv$  $W_{\,p\,q}(i_1\,j_1)$ as
\begin{center}
{\renewcommand\arraystretch{1.1}
\tabcolsep 14 pt
\doublerulesep 0pt
\begin{tabular}{c|rll} \hline
   {} &  $ p \backslash q $ &  ${1\over 2}$  &  $-{1\over 2}$     \\ \hline
  $V_{\,p\,q}(i_1\,j_1)$  &
                 $ {1\over 2} \hspace{3mm} $  
                 & $-\frac{K}{\sqrt{2}}E_{+i_1-\!j}  $
                 & $-\frac{K}{\sqrt{2}}E_{+i_1-\!j_1}$  \\        
        {} &     $ -{1\over 2} \hspace{3mm} $  
                 & $ \frac{K}{\sqrt{2}}E_{+i-\!j}    $
                 & $ \frac{K}{\sqrt{2}}E_{+i-\!j_1}  $    \\ \hline   
  $W_{\,p\,q}(i_1\,j_1)$  &
                 $ {1\over 2} \hspace{3mm} $  
                 & $ \frac{K}{\sqrt{2}}E_{-i+\!j_1}  $         
                 & $-\frac{K}{\sqrt{2}}E_{-i+\!j}    $   \\
        {} &     $ -{1\over 2} \hspace{3mm} $  
                 & $ \frac{K}{\sqrt{2}}E_{-i_1+\!j_1}$           
                 & $-\frac{K}{\sqrt{2}}E_{-i_1+\!j}  $   \\ \hline
\end{tabular}    }
\end{center}
where            
$
   i,\,j=1,\,3,\,5, \hspace{3mm} i<j, \hspace{3mm} i_1=i+1, \hspace{3mm}
   j_1=j+1,
$\\
$V_{\,p\,q}(1638)$ and $W_{\,p\,q}(1638)$ as
\begin{center}
{\renewcommand\arraystretch{1.1}
\tabcolsep 14 pt
\doublerulesep 0pt
\begin{tabular}{c|rll} \hline
   {} &  $ p \backslash q $ &  ${1\over 2}$  &  $-{1\over 2}$     \\ \hline
   $V_{\,p\,q}(1638)$  
                 & $ {1\over 2} \hspace{3mm} $  
                 & $-\frac{K}{\sqrt{2}}E_{\alpha_1}  $
                 & $-\frac{K}{\sqrt{2}}E_{\alpha_2}  $    \\    
       {}        & $ -{1\over 2} \hspace{3mm} $  
                 & $ \frac{K}{\sqrt{2}}E_{\alpha_3}  $
                 & $ \frac{K}{\sqrt{2}}E_{\alpha_4}  $   \\ \hline    
   $W_{\,p\,q}(1638)$  
                 & $ {1\over 2} \hspace{3mm} $  
                 & $ \frac{K}{\sqrt{2}}E_{-\alpha_4}  $
                 & $-\frac{K}{\sqrt{2}}E_{-\alpha_3}  $    \\    
       {}        & $ -{1\over 2} \hspace{3mm} $  
                 & $ \frac{K}{\sqrt{2}}E_{-\alpha_2}  $
                 & $ -\frac{K}{\sqrt{2}}E_{-\alpha_1}  $ \\ \hline
\end{tabular}    }
\end{center}
$V_{\,p\,q}(1458)$ and $W_{\,p\,q}(1458)$ as
\begin{center}
{\renewcommand\arraystretch{1.1}
\tabcolsep 14 pt
\doublerulesep 0pt
\begin{tabular}{c|rll} \hline
   {} &  $ p \backslash q $ &  ${1\over 2}$  &  $-{1\over 2}$     \\ \hline
   $V_{\,p\,q}(1458)$  
                 &  $ {1\over 2} \hspace{3mm} $ 
                 & $ \frac{K}{\sqrt{2}}E_{\beta_1}  $
                 & $ \frac{K}{\sqrt{2}}E_{\beta_2}  $    \\    
      {}         & $-{1\over 2} \hspace{3mm} $ 
                 & $-\frac{K}{\sqrt{2}}E_{\beta_3}  $
                 & $-\frac{K}{\sqrt{2}}E_{\beta_4}  $    \\ \hline   
   $W_{\,p\,q}(1458)$  
                 & $ {1\over 2} \hspace{3mm} $ 
                 & $-\frac{K}{\sqrt{2}}E_{-\beta_4} $          
                 & $ \frac{K}{\sqrt{2}}E_{-\beta_3} $      \\
       {}        & $-{1\over 2} \hspace{3mm} $ 
                 & $-\frac{K}{\sqrt{2}}E_{-\beta_2} $          
                 & $ \frac{K}{\sqrt{2}}E_{-\beta_1} $     \\ \hline
\end{tabular}      }
\end{center}        
$V_{\,p\,q}(3258)$ and $W_{\,p\,q}(3258)$ as
\begin{center}
{\renewcommand\arraystretch{1.1}
\tabcolsep 14 pt
\doublerulesep 0pt
\begin{tabular}{c|rll} \hline
   {} &  $ p \backslash q $ &  ${1\over 2}$  &  $-{1\over 2}$     \\ \hline
   $V_{\,p\,q}(3258)$  
                 & $ {1\over 2} \hspace{3mm} $ 
                 & $-\frac{K}{\sqrt{2}}E_{\epsilon_1} $ 
                 & $ \frac{K}{\sqrt{2}}E_{\epsilon_3} $    \\    
      {}         & $-{1\over 2} \hspace{3mm} $ 
                 & $-\frac{K}{\sqrt{2}}E_{\epsilon_2} $   
                 & $ \frac{K}{\sqrt{2}}E_{\epsilon_4} $      \\ \hline   
   $W_{\,p\,q}(3258)$  
                 & $ {1\over 2} \hspace{3mm} $ 
                 & $\frac{K}{\sqrt{2}}E_{-\epsilon_4}$                
                 & $ \frac{K}{\sqrt{2}}E_{-\epsilon_2}$       \\
       {}        & $-{1\over 2} \hspace{3mm} $ 
                 & $-\frac{K}{\sqrt{2}}E_{-\epsilon_3}$          
                 & $-\frac{K}{\sqrt{2}}E_{-\epsilon_1}$       \\ \hline
\end{tabular}      }
\end{center}        
$U_{\;p\:q\:p'\:q'}^{{1\over 2}{1 \over 2}\,{1 \over 2}\:{1 \over 2}}
(2468) \equiv U_{p\: q\: p'\: q'}(2468)$  as
\begin{center}
{\renewcommand\arraystretch{1.1}
\tabcolsep 12 pt
\doublerulesep 0pt
\begin{tabular}{rllll}\hline
       $  p\,q \backslash p' \, q' $  
       &  ${1\over 2}\; {1\over 2}$  & ${1\over 2}\; -{1\over 2}$ 
       &  $-{1\over 2}\; {1\over 2}$ & $-{1\over 2}\; -{1\over 2}$
       \\[1.5mm] \hline
${1\over 2}\hspace{5mm}{1\over 2} \hspace{4mm}  $  
                           & $-\frac{K}{\sqrt{2}}E_{\lambda_1}  $
                           & $ \frac{K}{\sqrt{2}}E_{\lambda_2}  $  
                           & $-\frac{K}{\sqrt{2}}E_{\lambda_3}  $
                           & $-\frac{K}{\sqrt{2}}E_{\lambda_4}  $\\
${1\over 2}\; -\!{1\over 2}  \hspace{4mm} $ 
                           & $ \frac{K}{\sqrt{2}}E_{\lambda_5}$ 
                           & $-\frac{K}{\sqrt{2}}E_{\lambda_6}  $  
                           & $ \frac{K}{\sqrt{2}}E_{\lambda_7}  $
                           & $ \frac{K}{\sqrt{2}}E_{\lambda_8}  $\\
$-{1\over 2}\hspace{5mm}{1\over 2} \hspace{4mm}  $
                           & $ \frac{K}{\sqrt{2}}E_{-\lambda_8}  $
                           & $-\frac{K}{\sqrt{2}}E_{-\lambda_7}  $  
                           & $ \frac{K}{\sqrt{2}}E_{-\lambda_6}  $
                           & $ \frac{K}{\sqrt{2}}E_{-\lambda_5}  $\\
$-{1\over 2}\;-\!{1\over 2} \hspace{4mm} $ 
                           & $ \frac{K}{\sqrt{2}}E_{-\lambda_4}  $
                           & $-\frac{K}{\sqrt{2}}E_{-\lambda_3}  $  
                           & $ \frac{K}{\sqrt{2}}E_{-\lambda_2}  $
                           & $ \frac{K}{\sqrt{2}}E_{-\lambda_1}  $
\\ \hline  
\end{tabular}    }
\end{center}
The number of the above operators is  
$ 3\times 3 + 2 + 3 + 6\times 8 + 16 = 78$, it is equal to the order
of $E_6$. Hence, these operators form the irreducible tensor basis
of $E_6$.

\par
It is not difficult to find
\begin{eqnarray*} 
V_{\,-p\,-q} = (-)^{1+p+q} \, W_{\,p\,q}^{\dagger},
\end{eqnarray*}
\begin{eqnarray*} 
W_{\,-p\,-q} = (-)^{1+p+q} \, V_{\,p\,q}^{\dagger};
\end{eqnarray*}
\begin{eqnarray*} 
U_{\,-p\,-q\,-p'\,-q'}
     = (-)^{1+p+q+p'+q'}\, U_{\,p\,q\,p'\,q'}^{\dagger}.
\end{eqnarray*}

\par
By direct calculations, we can obtain the commutation relations
satisfied by the irreducible tensor basis of $E_6$: 
\par
1) {\bf J}(2), {\bf J}(4), {\bf J}$(6)$ and {\bf J}$(8)$ are
   the mutually commuting angular momentum operators,
   and satisfy commutation relations (\ref{angular-d}).

\par
2) $A(1)$, $A(3)$ and $A(5)$ (only two of them are independent)
are the mutually commuting scalar operators, hence they, together 
with ${\bf J}(1)$, ${\bf J}(3)$ and ${\bf J}(5)$, satisfy
commutation relations (\ref{scalar-d}).

\par
3) Both ${\bf V}^{{1\over 2}{1\over 2}}(i\,j_1\,k\,l_1)$ and
${\bf W}^{{1\over 2}{1\over 2}}(i\,j_1\,k\,l_1)$ are
the 2-fold irreducible tensor operators, hence they, 
together with {\bf J}$(j_1)$ and {\bf J}$(l_1)$, satisfy 
commutation relations (\ref{mtensr-d}). 

4) The nonzero commutation relations satisfied by sixty-four 
components of ${\bf V}^{{1\over 2}{1\over 2}}(i\,j_1\,k\,l_1)$ and
${\bf W}^{{1\over 2}{1\over 2}}(i\,j_1\,k\,l_1)$ and two scalar operators are
\begin{eqnarray}
  \begin{array}{l}
     [ A(i), \, V_{\,p\,q}(i\,j_1\,k\,l_1)]
            =  V_{\,p\,q}(i\,j_1\,k\,l_1),        \cr
     [ A(i), \, W_{\,p\,q}(i\,j_1\,k\,l_1)]
            = - W_{\,p\,q}(i\,j_1\,k\,l_1),       \cr
     [ A(k), \, V_{\,p\,q}(i\,j_1\,k\,l_1)]
            = -V_{\,p\,q}(i\,j_1\,k\,l_1),        \cr
     [ A(k), \, W_{p\,q}(i\,j_1\,k\,l_1)]
            =  W_{\,p\,q}(i\,j_1\,k\,l_1);        \cr
 \left[ {\bf V}^{{1\over 2}{1\over 2}}(i\,j_1\,k\,l_1) \,
        {\bf W}^{{1\over 2}{1\over 2}}(i\,j_1\,k\,l_1) 
 \right]^{\:\;1\;\;0}_{0\mu 00} = J_{\mu}(j_1),                           \cr
 \left[ {\bf V}^{{1\over 2}{1\over 2}}(i\,j_1\,k\,l_1)\,
        {\bf W}^{{1\over 2}{1\over 2}}(i\,j_1\,k\,l_1)
 \right]^{\;\;0\;\:1}_{000\mu } = J_{\mu }(l_1) ,                         \cr
 \mu =-1,\,0,\,1;                                                        \cr
 \left[ {\bf V}^{{1\over 2}{1\over 2}}(i\,j_1\,k\,l_1)\,
        {\bf W}^{{1\over 2}{1\over 2}}(i\,j_1\,k\,l_1)
 \right] ^{\;\:0\;\:0}_{0000} = {1\over 2} \, \left[ A(i)- A(k) \right] ;  \cr
 \left\{ {\bf V}^{{1\over 2}{1\over 2}}(i_1\,j_1) \,
         {\bf V}^{{1\over 2}{1\over 2}}(j_1\,k_1)
 \right\}^{\hspace{2.2mm}{1 \over 2} \hspace{2mm}0 \hspace{5mm}{1 \over 2}}
         _{\,1\,p\,0\,0\,-1\,q} 
     =  V^{{1\over 2}{1\over 2}}_{\,p\,q}(i_1\, k_1) ,                       \cr
 \left\{ {\bf W}^{{1\over 2}{1\over 2}}(i_1\, j_1) \, 
         {\bf W}^{{1\over 2}{1\over 2}}(j_1\, k_1) 
 \right\}^{\hspace{4.2mm}{1 \over 2} \hspace{2.5mm} 0 \hspace{2.2mm}{1 \over 2}}
         _{\,-1\,p\,0\,0\,1\,q} 
     =  W^{{1\over 2}{1\over 2}}_{\,p\,q}(i_1\,k_1) .                             
 \end{array}
\label{12}
\end{eqnarray}
We can conclude from the former four equations in Eq. (\ref{12}) that 
${\bf V}^{{1\over 2}{1\over 2}}(i\,j_1\,k\,l_1)$ raise and lower 
the eigenvalues of $A(i)$ and $A(k)$ by $1$ respectively, while
${\bf W}^{{1\over 2}{1\over 2}}(i\,j_1\,k\,l_1)$ lower and raise
the eigenvalues of $A(i)$ and $A(k)$ by $1$ respectively.

\par
4) ${\bf U}^{{1\over 2}{1\over 2}{1\over 2}{1\over 2}}(2468)$ is
a 4-fold irreducible tensor operator, hence it, together with {\bf J}(2), 
{\bf J}(4), {\bf J}$(6)$ and {\bf J}$(8)$ satisfies the commutation relations
(\ref{mtensr-d}). The nonzero commutation relations between components of 
${\bf U}^{{1\over 2}{1\over 2}{1\over 2}{1\over 2}}(2468)$ are the last four
equations in Eq. (\ref{f4-cr}).

\par
5) The other nonzero commutation relations satisfied by these irreducible tensor
operators are
{\small
\begin{eqnarray}
  \begin{array}{ll}
   [V(1638),\; W(1458)]= -\sqrt{1 \over 2}\, W(3456),       &
   [W(1638),\; V(1458)]= +\sqrt{1 \over 2}\, V(3456),       \cr
   [V(3456),\; W(1638)]= -\sqrt{1 \over 2}\, W(1458),       &
   [W(3456),\; V(1638)]= +\sqrt{1 \over 2}\, V(1458),       \cr
   [V(3456),\; W(1458)]= +\sqrt{1 \over 2}\, W(1638),       &
   [W(3456),\; V(1458)]= -\sqrt{1 \over 2}\, V(1638);       \cr
   [V(1638),\; W(3258)]= +\sqrt{1 \over 2}\, W(1256),       &
   [W(1638),\; V(3258)]= -\sqrt{1 \over 2}\, V(1256),       \cr
   [V(1256),\; W(1638)]= +\sqrt{1 \over 2}\, W(3258),       &
   [W(1256),\; V(1638)]= -\sqrt{1 \over 2}\, V(3258),       \cr
   [V(1256),\; W(3258)]= -\sqrt{1 \over 2}\, W(1458),       &
   [W(1256),\; V(3258)]= +\sqrt{1 \over 2}\, V(1458);       \cr
   [V(1458),\; W(3258)]= -\sqrt{1 \over 2}\, W(1234),       &
   [W(1458),\; V(3258)]= +\sqrt{1 \over 2}\, V(1234),       \cr
   [V(1234),\; W(1458)]= -\sqrt{1 \over 2}\, W(3258),       &
   [W(1234),\; V(1458)]= +\sqrt{1 \over 2}\, V(3258),       \cr
   [V(1234),\; W(3258)]= +\sqrt{1 \over 2}\, W(1458),       &   
   [W(1234),\; V(3258)]= -\sqrt{1 \over 2}\, V(1458);      \cr  
   [V(1234),\; W(1638)]= +\sqrt{1 \over 2}\, U(2468),       &
   [W(1234),\; V(1638)]= -\sqrt{1 \over 2}\, U(2468),       \cr
   [V(1234),\; U(2468)]= -\sqrt{1 \over 2}\, V(1638),       &
   [W(1234),\; U(2468)]= +\sqrt{1 \over 2}\, W(1638),       \cr
   [V(1638),\; U(2468)]= +\sqrt{1 \over 2}\, V(1234),       &
   [W(1638),\; U(2468)]= -\sqrt{1 \over 2}\, W(1234);       \cr
   [V(1256),\; W(1458)]= +\sqrt{1 \over 2}\, U(2468),       &
   [W(1256),\; V(1458)]= -\sqrt{1 \over 2}\, U(2468),       \cr
   [V(1256),\; U(2468)]= -\sqrt{1 \over 2}\, V(1458),       &
   [W(1256),\; U(2468)]= +\sqrt{1 \over 2}\, W(1458),       \cr
   [V(1458),\; U(2468)]= +\sqrt{1 \over 2}\, V(1256),       &
   [W(1458),\; U(2468)]= -\sqrt{1 \over 2}\, W(1256);       \cr
   [V(3456),\; W(3258)]= +\sqrt{1 \over 2}\, U(2468),       &
   [W(3456),\; V(3258)]= -\sqrt{1 \over 2}\, U(2468),       \cr
   [V(3456),\; U(2468)]= -\sqrt{1 \over 2}\, V(3258),       &
   [W(3456),\; U(2468)]= +\sqrt{1 \over 2}\, W(3258),       \cr
   [V(3258),\; U(2468)]= +\sqrt{1 \over 2}\, V(3456),       &  
   [W(3258),\; U(2468)]= -\sqrt{1 \over 2}\, W(3456).        
  \end{array}
\label{e6-cr2}
\end{eqnarray}
}
In Eq. (\ref{e6-cr2}),
we have utilized the simple expressions, for example,  
$$
  [ V(1638),\; W(1458) ]= -\sqrt{1 \over 2}\, W(3456)
$$
means
$$
  \left[ V^{{1\over 2}{1\over 2}}_{\,q\,q'}(1638), \;
         W^{{1\over 2}\hspace{2mm} {1\over 2}}_{\,p\,-q'}(1458) \right] 
    = -(2q')\sqrt{1 \over 2}\, W^{{1\over 2}{1\over 2}}_{\,p\,q}(3456),
$$
and so forth.

\par
We can also find from the above commutation relations that 
the Cartan generators of $E_6$ in the irreducible tensor basis 
are $\{ A(1)$, $A(3)$, $J_0(2)$, $J_0(4)$, $J_0(6)$, $J_0(8) \}$.

\section{Conclusions}
In this paper, we obtain the irreducible tensor bases of exceptional 
Lie algebras $G_2$, $F_4$ and $E_6$ by grouping their Cartan-Weyl bases
according to the respective chains $G_2$ $\supset$ SO(3) $\otimes$ SO(3),
$F_4$ $\supset$ SO(3) $\otimes$ SO(3) $\otimes$ SO(3) $\otimes$ SO(3)
and $E_6$ $\supset$ SO(3) $\otimes$ SO(3) $\otimes$ SO(3) $\otimes$ SO(3). 
The irreducible tensor basis of $G_2$ is made up of
two mutually commuting angular momentum operators and
one 2-fold irreducible tensor operator of rank
(${1\over 2}{3\over 2})$.
The irreducible tensor basis of $F_4$ is made up of four mutually
commuting angular momentum operators, six 2-fold irreducible tensor
operators of rank (${1\over 2}{1\over 2}$) and one 4-fold irreducible
tensor operator of rank (${1\over 2}{1\over 2}{1\over 2}{1\over 2}$).
The irreducible tensor basis of $E_6$ is made up of two independent
and mutually commuting scalar operators, four mutually commuting
angular momentum operators, twelve 2-fold irreducible tensor operators
of rank (${1\over 2}{1\over 2}$) and one 4-fold irreducible tensor
operator of rank (${1\over 2}{1\over 2}{1\over 2}{1\over 2}$).
However, it is worth reminding readers to note that, 
in the process of constructing the angular momentum operators and 
the (multi-fold) irreducible tensor operators within the irreducible 
tensor basis, the structure constants of the Cartan-Weyl basis can not 
be taken arbitrarily even if they obey the symmetries and identities
[In the irreducible tensor bases of Racah's type for some exceptional 
Lie algebras, one or two free parameters exist. 
\cite{jbm,jeugt92,berghe,bmj,mbj}]
The explicit commutation relations satisfied by the irreducible tensor 
bases of $G_2$, $F_4$ and $E_6$ are gained as well respectively.
Thus, by means of the similar method used in Refs. \cite{sh2,hs2,hls,sr1,shzr} 
(especially, the Wigner-Eckart theorem \cite{edmonds,wy,bl,wigner}), 
the problems of irreducible representations of exceptional Lie algebras 
may be solved. They are being studied.
The irreducible tensor bases of exceptional Lie algebras $E_7$ and $E_8$
will be discussed by considering the suitable chains in a subsequent paper.

\section*{Acknowledgments}
The project supported by National Natural Science Foundation of China
(19905005), Major State Basic Research Development Programs (G2000077400 and 
G2000077604) and Tsinghua Natural Science Foundation (985 Program).

\newpage

%%%%%%%%%%%%%%%%%%%%%%%%%%%%%%%%%%  Caption    %%%%%%%%%%%%%%%%%%%%%%%%%%%

\begin{center}
{\Large\bf Captions}
\end{center}

\vspace{1cm}

TABLE 1  \hspace{2mm} $G_{xy}$  of $F_4$  

\vspace{1cm}

TABLE 2 \hspace{2mm} $S_{xy}$  of $E_6$ 

\vspace{1cm}

FIGURE 1 \hspace{2mm} The root diagram corresponding to
the irreducible tensor basis of $G_2$

\newpage

\begin{center}
{\bf TABLE 1} \hspace{2mm}   $G_{xy}$  of $F_4$ \\ 
{\small\doublerulesep 0pt
\def\arraystretch{1.1} 
\begin{tabular}{cccc}
$     \begin{array}{rll}  \hline
      x \backslash y  &  \alpha_2  & -\alpha_2        \\ \hline
      \alpha_1  &    +1           &   +1              \\ \hline
     \end{array}      
$
& \hfill  
$     \begin{array}{rll} \hline
      x \backslash y  &  \beta_2   & -\beta_2     \\ \hline
      \beta_1   &    +1        &   +1           \\ \hline
      \end{array}     $
   & \hfill   
$     \begin{array}{rll} \hline
      x \backslash y   &  \gamma_2  & -\gamma_2     \\ \hline
      \gamma_1  &    +1        &   +1               \\ \hline
     \end{array}   $
   & \hfill  
$     \begin{array}{rll} \hline
      x \backslash y   &  \epsilon_2   & -\epsilon_2   \\ \hline
      \epsilon_1   &    +1           &   +1           \\ \hline
      \end{array}  $ 
\end{tabular} }
\vspace{8mm}

{\doublerulesep 0pt
\def\arraystretch{1} 
\begin{tabular}{ccc}
$     \begin{array}{rcc}  \hline
      x \backslash y  
                &  -\beta_2          & -\beta_1           \\ \hline
      \alpha_1  &  +\sqrt{1\over 2}  &  -\sqrt{1\over 2}  \\
      \alpha_2  &  +\sqrt{1\over 2}  &  +\sqrt{1\over 2}   \\ \hline
     \end{array}     $
   & \hfill   
$     \begin{array}{rcc}  \hline
      x \backslash y  
                &  -\epsilon_2       &   -\epsilon_1        \\ \hline
      \gamma_1  &  -\sqrt{1\over 2}  &   +\sqrt{1\over 2}   \\
      \gamma_2  &  -\sqrt{1\over 2}  &   -\sqrt{1\over 2}   \\ \hline
     \end{array}    $
   & \hfill   
$     \begin{array}{rcc}  \hline
      x \backslash y  
                &   \gamma_2         & -\gamma_1          \\ \hline
      \alpha_1  &  +\sqrt{1\over 2}  &  -\sqrt{1\over 2}  \\
     -\alpha_2  &  +\sqrt{1\over 2}  &  +\sqrt{1\over 2}   \\ \hline
     \end{array}     $
\end{tabular} }
\vspace{8mm}

{\doublerulesep 0pt
\def\arraystretch{1} 
\begin{tabular}{ccc}
$  \begin{array}{rcccc}  \hline
      x \backslash y  
               &    \epsilon_2      &   -\epsilon_1       \\ \hline
      \beta_1  &  +\sqrt{1\over 2}  &   -\sqrt{1\over 2}  \\
     -\beta_2  &  +\sqrt{1\over 2}  &    +\sqrt{1\over 2}   \\ \hline
     \end{array}      $
 & \hfill   
$  \begin{array}{rcccc}  \hline
      x \backslash y  
             &  \epsilon_1   &  -\epsilon_1  &  \epsilon_2  & -\epsilon_2 \\ \hline
   \alpha_1  &  +1           &      -1       &      +1      &        +1   \\
  -\alpha_2  &  +1           &      +1       &      +1      &        -1    \\
  \hline
  \end{array}   $
 & \hfill  
$  \begin{array}{rcccc}  \hline
      x \backslash y  
              &  \gamma_1   &  -\gamma_1  &  \gamma_2  &  -\gamma_2 \\ \hline
     \beta_1  &  +1         &       +1     &     -1     &        +1    \\
    -\beta_1  &  +1         &       -1     &     -1     &        -1    \\
  \hline
  \end{array}       $
\end{tabular}  }
\end{center}

\newpage

\begin{center}
{\bf TABLE 2} \hspace{2mm} $S_{xy}$ of $E_6$ \\ 
{\small\doublerulesep 0pt
\def\arraystretch{1} 
\begin{tabular}{cccc}
$     \begin{array}{rll}  \hline
      x \backslash y   & -\alpha_3   & -\alpha_2  \\ \hline
      \alpha_1  &  +1    &  -1   \\
      \alpha_4  &  -1    &  +1   \\ \hline
     \end{array}   $
   & \hfill   
$     \begin{array}{rll} \hline
      x \backslash y   & -\alpha_3   & -\alpha_1  \\ \hline
      \beta_1  &  +1         &  +1   \\
      \beta_3  &  +1         &  +1   \\ \hline
     \end{array}    $
   & \hfill   
$     \begin{array}{rll} \hline
      x \backslash y   & -\alpha_4   & -\alpha_2  \\ \hline
      \beta_2  &  +1         &  +1   \\
      \beta_4  &  +1         &  +1   \\ \hline
     \end{array}    $
   & \hfill   
$     \begin{array}{rll}  \hline
      x \backslash y   & -\beta_3   & -\beta_2  \\ \hline
      \beta_1  &  +1        &  -1   \\
      \beta_4  &  -1        &  +1   \\ \hline
     \end{array}   $
\end{tabular}   }
\end{center}
\vspace{4mm}

\begin{center}
{\doublerulesep 0pt
\def\arraystretch{1} 
\begin{tabular}{cccc}
$     \begin{array}{rll} \hline
      x \backslash y   & -\epsilon_3   & -\epsilon_2  \\ \hline
      \epsilon_1  &  +1           &  -1   \\
      \epsilon_4  &  -1           &  +1   \\ \hline
     \end{array}  $
   & \hfill  
$     \begin{array}{rll} \hline
      x \backslash y   & -\lambda_3   & -\lambda_2  \\ \hline
      \lambda_1  &  -1          &  +1   \\
      \lambda_4  &  +1          &  -1   \\ \hline
     \end{array}   $
   & \hfill   
$     \begin{array}{rll} \hline
      x \backslash y    & \lambda_8  & -\lambda_5  \\ \hline
      \lambda_1  &  +1        &  +1   \\
     -\lambda_4  &  -1        &  -1   \\ \hline
     \end{array}  $
   & \hfill   
$     \begin{array}{rll} \hline
      x \backslash y   & \lambda_7   & -\lambda_6  \\ \hline
      \lambda_2  &  +1         &  +1   \\
     -\lambda_3  &  -1         &  -1   \\ \hline
     \end{array}   $
\end{tabular} }    
\end{center}
\vspace{4mm}

\begin{center}
{\doublerulesep 0pt
\def\arraystretch{1} 
\begin{tabular}{ccc}
$     \begin{array}{rll} \hline
      x \backslash y   & -\lambda_7   & -\lambda_6  \\ \hline
      \lambda_5  &  -1          &  +1   \\
      \lambda_8  &  -1          &  +1   \\ \hline
     \end{array}  $
   & \hfill   
$     \begin{array}{rllll} \hline
      x \backslash y  
     &  -\lambda_5   &   -\lambda_1  &  \lambda_4  &  \lambda_8 \\ \hline
   \alpha_1  &  -1           &      -1       &    -1       &       +1     \\
  -\alpha_4  &  +1           &      -1       &    -1       &       -1    \\ \hline 
  \end{array}        $
 & \hfill
$     \begin{array}{rllll} \hline
      x \backslash y  
               &  -\lambda_6  &  -\lambda_2  &  \lambda_3  &   \lambda_7 \\ \hline
     \alpha_2  &  +1          &      +1     &     +1     &       -1    \\
    -\alpha_3  &  +1          &      -1     &     -1     &       -1    \\ \hline
  \end{array}       $
\end{tabular}  }
\end{center}
\vspace{4mm}

\begin{center}
{\doublerulesep 0pt
\def\arraystretch{1} 
\begin{tabular}{cc}
$     \begin{array}{rllll} \hline
      x \backslash y  
          &  -\lambda_3   &   -\lambda_1  &  \lambda_6  &  \lambda_8 \\ \hline
  \beta_1 &  -1           &      +1       &    +1       &       -1     \\
 -\beta_4 &  +1           &      +1       &    +1       &       +1     \\ \hline
  \end{array}   $
 & \hfill
$     \begin{array}{rllll} \hline
      x \backslash y  
             &  -\lambda_4  &  -\lambda_2  &  \lambda_5  &   \lambda_7 \\ \hline
   \beta_2   &  -1          &      -1     &     +1     &       +1    \\
  -\beta_3   &  -1          &      +1     &     -1     &       +    \\ \hline
  \end{array}    $
\end{tabular}   }
\end{center}
\vspace{4mm}

\begin{center}
{\doublerulesep 0pt
\def\arraystretch{1} 
\begin{tabular}{cc}
$     \begin{array}{rllll} \hline
      x \backslash y  &  \alpha_2   &   \alpha_4  &  -\beta_3  &  -\beta_1   \\ \hline
  \epsilon_1 &  -1          &      +1     &    +1      &       -1     \\
  \epsilon_3 &  -1          &      +1     &    -1      &       +1     \\ \hline
  \end{array}      $
 & \hfill
$     \begin{array}{rllll} \hline
      x \backslash y 
                & \alpha_1   & \alpha_3   & -\beta_4   & -\beta_2  \\ \hline
   \epsilon_2   &  +1        &      -1     &     +1     &       -1    \\
   \epsilon_4   &  +1        &      -1     &     -1     &       +1    \\ \hline
  \end{array}      $
\end{tabular}  }
\end{center}
\vspace{4mm}

\begin{center}
{\doublerulesep 0pt
\def\arraystretch{1} 
\begin{tabular}{cc}
$     \begin{array}{rllll} \hline
      x \backslash y
               &  -\lambda_7   &   -\lambda_5  &  -\lambda_3  &  -\lambda_1 
               \\ \hline
   \epsilon_1  &  -1           &      +1       &    +1       &       -1     \\
  -\epsilon_4  &  -1           &      -1       &    -1       &       -1     \\ \hline       
  \end{array}             $
 & \hfill
$     \begin{array}{rllll} \hline
      x \backslash y
               &  -\lambda_8  &  -\lambda_6  &  \lambda_4  &   \lambda_2 \\ \hline
   \epsilon_2  &  -1          &      -1     &     +1     &       +1    \\
  -\epsilon_3  &  +1          &      -1     &     +1     &       -1    \\ \hline
  \end{array}     $
\end{tabular}     }
\end{center}
\end{document}